\documentclass[english,12pt]{article}
\usepackage{array}
\usepackage{graphicx}
\usepackage{amssymb}
\usepackage{amsmath}
\usepackage{multirow}
\usepackage{prettyref}
\usepackage{babel}
\usepackage{units}
\usepackage[latin1]{inputenc}
\usepackage{amsfonts}
\usepackage{amssymb}
\usepackage{babel}
\usepackage{subcaption}
\usepackage[dvipsnames,usenames]{color}
\usepackage{cite}
\def\@fmsl@sh#1#2#3{\m@th\ooalign{$\hfil#1\mkern#2/\hfil$\crcr$#1#3$}}
 \def\eq#1\en{\begin{equation}#1\end{equation}}
\def\s[#1,#2]{[#1\stackrel{\star}{,}#2]}
\def\sx[#1,#2]{[#1\stackrel{\star_{x}}{,}#2]}

\def\beq{\begin{equation}}
\def\eeq{\end{equation}}
\newcommand{\pro}[2]{\mbox{$\langle\, #1 \mid #2\,\rangle$}}
\newcommand{\expec}[1]{\mbox{$\langle\, #1\,\rangle$}}

\renewcommand{\d}{\mbox{${\rm d}$}} 
\newcommand{\lp}{\ell_{\rm p}}
\newcommand{\mpl}{m_{\rm p}}
\newcommand{\gn}{G_{\rm N}}

\newcommand{\bra}[1]{\langle\,#1\,|}
\newcommand{\ket}[1]{|\,#1\,\rangle}
\def\gsim{\mathrel{\rlap{\lower4pt\hbox{\hskip1pt$\sim$}}\raise1pt\hbox{$>$}}}      
\begin{document}
\makeatletter
\def\fmslash{\@ifnextchar[{\fmsl@sh}{\fmsl@sh[0mu]}}
\def\fmsl@sh[#1]#2{%
  \mathchoice
    {\@fmsl@sh\displaystyle{#1}{#2}}%
    {\@fmsl@sh\textstyle{#1}{#2}}%
    {\@fmsl@sh\scriptstyle{#1}{#2}}%
    {\@fmsl@sh\scriptscriptstyle{#1}{#2}}}
\def\@fmsl@sh#1#2#3{\m@th\ooalign{$\hfil#1\mkern#2/\hfil$\crcr$#1#3$}}
\makeatother

\thispagestyle{empty}
\begin{titlepage}
\begin{center}
 \Large {\bf Revisiting the minimum length in the Schwinger-Keldysh formalism}
\end{center}
\vspace{0.2cm}
\begin{center}
{{Roberto~Casadio$^{ab}$}\footnote{casadio@bo.infn.it}
and
{Iber\^e Kuntz$^{ab}$}\footnote{kuntz@bo.infn.it}}
 \end{center}
\begin{center}
{\sl
$^a$
Dipartimento di Fisica e Astronomia, Universit\`a di Bologna,
\\
via Irnerio~46, I-40126 Bologna, Italy}
\\
$ $
\\
{\sl 
$^b$
I.N.F.N., Sezione di Bologna, IS - FLAG
\\
via B.~Pichat~6/2, I-40127 Bologna, Italy
}
\end{center}
\vspace{2cm}
\begin{abstract}
The existence of a minimum length in quantum gravity is investigated by computing
the in-in expectation value of the proper distance in the Schwinger-Keldysh formalism.
No minimum geometrical length is found for arbitrary gravitational theories
to all orders in perturbation theory.
Using non-perturbative techniques, we also show that neither the conformal sector
of general relativity nor higher-derivative gravity features a minimum length.
A minimum length scale, on the other hand, seems to always be present when one
considers in-out amplitudes, from which one could extract the energy scale of scattering
processes.
\end{abstract}  
\end{titlepage}
%
%
%
%
\newpage
\section{Introduction}
\label{S:intro}
As we live in a world where all of our daily observations take place at scales such as the meter,
the second and the kilogram, it is not easy for modern human minds to grasp the possibility
that there exists fundamental upper or lower bounds on physical quantities that could otherwise
become evident at much smaller or larger scales.
Our experience and the convenience of describing it with continuum mathematics therefore
make us think that it is natural for physical quantities to admit an infinite range of possible values.
In fact, nothing in classical mechanics forbids us from speeding to infinity or dismantling the
spacetime into infinitesimally small distances.
Yet it is a fact of nature that there exists a limiting speed, which special relativity incorporates
and allows us to describe its kinematical consequences.  
Naturally, this raises the similar question of whether it is possible to probe decreasingly
small lengths or if there is a limiting factor that keep us from accessing some fundamental
length scales.
\par
The notion of a minimum length (see Ref.~\cite{Hossenfelder:2012jw} for a in-depth review) dates back
to the early days of quantum field theory, when physicists were desperately attempting to get rid of the
troubling ultraviolet divergences, but it soon became unattractive with the advent of the more sophisticated
methods of renormalization.
It only regained notoriety with the increasing interest in trans-Planckian effects.
Currently, many models of quantum gravity exhibit some notion of minimum length, including string theory,
loop quantum gravity, asymptotically safe gravity and the conformal sector of general relativity.
However, some works have established the possibility of a minimum geometrical
length by employing the standard Feynman path integral for the calculation of time-ordered
in-out amplitudes~\cite{Padmanabhan:1985jq}.
These amplitudes are the correct ingredients for obtaining $S$-matrix elements from the LSZ formula,
but are otherwise acausal and complex, being subjected to Feynman boundary conditions.
Taken literally, an observable minimum length in quantum gravity should be real to all loop orders
and bare the statistical properties of an expectation value.
In this respect, it is therefore very important to distinguish between the use of in-out amplitudes and
in-in amplitudes, the latter being the objects which admit a proper statistical interpretation.
These requirements lead us to study the minimum length using the in-in expectation value,
which can be obtained in the Schwinger-Keldysh path integral formalism~\cite{Keldysh:1964ud}
and whose evolution is subjected to retarded boundary conditions~\cite{Jordan:1986ug}.
\par
The main goal of this paper is to investigate the distinct properties of the in-in proper distance,
which can be directly interpreted as a geometrical length, and the in-out proper ``length'',
which cannot be interpreted as a physical distance but sets the length scale of the underlying
scattering process.
As we will see, the former vanishes quite generally at the coincidence limit, suggesting that
a geometrical minimum length is most likely absent.
On the other hand, when the latter is evaluated at the coincidence limit, it acquires a finite
value of the order of the Planck scale under very general assumptions,
indicating that a minimum length \textit{scale} is very likely to exist.
The implication of these results is that nothing prevents one from going through vanishingly
small distances {\em in principle\/}, but scattering experiments cannot reliably distinguish
between events taking place at the Planck scale, since any two processes differing only
at trans-Planckian scales would produce the same scattering amplitudes.
\par
This paper is organized as follows:
in Section~\ref{inin}, we briefly review some aspects of the Schwinger-Keldysh formalism
used for the calculation of in-in amplitudes;
in Section~\ref{general}, we show that a minimum length cannot exist to second order
in the metric perturbation for any metric theory of gravity whose gravitational propagator
can be written as the sum of partial fractions of the form $(q^2 - m^2)^{-1}$,
but a minimum length scale is always present.
The absence of interactions allows the extension of this result to all orders in pertubation theory,
although interacting theories would require the evaluation of higher-order amplitudes;
Section~\ref{hdgravity} is devoted to the study of a minimum length in higher-derivative gravity.
Without resorting to perturbation theory, we show that higher-derivative gravity does not exhibit
any obstruction to the continuous shrinkage of the quantum proper length to zero;
in Section~\ref{conf}, we revisit the conformal degree of freedom in gravity,
which had previously been shown to yield a ground-state length in the in-out approach.
The Schwinger-Keldysh formalism allows us to show that the minimum length is again
absent in this theory;
we finally draw our conclusions and briefly compare with other approaches
in Section~\ref{conc}.
\section{Schwinger-Keldysh formalism}
\label{inin}
Before elaborating on the minimum length, we need to clarify an important point
that has been largely ignored in the literature.
In all calculations of the expectation value of the proper length
$\expec{\d s^2}$, the in-out formalism has been implicitly employed
with no proper justification, which makes $\expec{\d s^2}$ a short-hand
notation for $\bra{0_\text{out}}\d s^2\ket{0_\text{in}}$,
namely some sort of transition amplitude from an in-vacuum state
to an out-vacuum state.
This is the standard kind of amplitude obtained from functional derivatives of the
generating functional $Z[J]$ which results from Feynman path integrals and
satisfies Feynman boundary conditions.
Transition amplitudes are in general acausal and complex (even for Hermitian operators)
distributions, thus they cannot make up the list of observables of a quantum field theory.
This is usually not an issue because they only show up in intermediate steps of the
calculation of $S$-matrix components, eventually yielding cross sections, which are the ultimate
object of interest in scattering experiments.
\par
Although the in-out formalism is the standard approach for the calculation of scattering amplitudes, 
its use obscures the physical interpretation of the quantum proper length.
For the above reasons, the fact that $\ket{0_\text{in}}\neq \ket{0_\text{out}}$ makes
it impossible to interpret $\bra{0_\text{out}}\d s^2\ket{0_\text{in}}$ as an expectation
value or to attribute to it any statistical meaning.
It appears hard to accept that a length which is neither real nor respects causality can bare any
physical reality. 
In order to talk of a minimum length, we need to calculate
$\bra{0_\text{in}}\d s^2\ket{0_\text{in}}$ instead, namely the quantum proper length
evaluated on one and the same quantum state $\ket{0_\text{in}}$.
It is important to remark that the in-in mean field $\bra{0_\text{in}}\phi\ket{0_\text{in}}$ not only
is real for Hermitian operators $\phi$, it also evolve causally, which is particularly important for
time-dependent settings in which ones does not know, or is not interested in, the final state
$\ket{0_\text{out}}$ of the system.
\par
The calculation of in-in amplitudes does not follow directly from the usual Feynman path integral,
but it can be performed using the slightly different Schwinger-Keldysh formalism
(or closed-time path integral)~\cite{Keldysh:1964ud,Jordan:1986ug}.
The idea is to double each degree of freedom $\phi$ and commonly denote the two peers with
$\phi_+$ and $\phi_-$.
The field $\phi_+$ is generated by an external source $J_+$ and is responsible for the transition
between $\ket{0_\text{in}}$ and an intermediate state $\ket{\Sigma_\alpha}$
belonging to a future Cauchy surface $\Sigma$, while $\phi_-$ is generated by $J_-$ and
takes care of the transition from $\ket{\Sigma_\alpha}$ back to $\ket{0_\text{in}}$.
Assuming $\{\ket{\Sigma_\alpha}\}$ form a complete set of states, the functional generator
of connected in-in correlation functions is then obtained by summing over all possible intermediate
states $\ket{\Sigma_\alpha}$, to wit
\beq
e^{i\,W[J_+, J_-]}
=
\sum_\alpha
\pro{0_\text{in}}{\Sigma_\alpha}_{J_-}
\pro{\Sigma_\alpha}{0_\text{in}}_{J_+}
\ .
\label{eq:gen}
\eeq
If we further assume that $\{\ket{\Sigma_\alpha}\}$ are eigenstates of $\phi$ on $\Sigma$,
we can write Eq.~\eqref{eq:gen} in terms of Feynman path integrals as
\beq
e^{i\,W[J_+,J_-]}
=
\int\mathcal D\phi_+\, \mathcal D\phi_-\,
e^{\frac{i}{\hbar}\left\{S[\phi_+] + S[\phi_-] + J_+\, \phi_+ - J_-\, \phi_- \right\}}
\ ,
\eeq
where the integration variables are subjected to vacuum boundary conditions in the remote past
(corresponding to the state $\ket{0_\text{in}}$) and $\phi_+=\phi_-$ on $\Sigma$.
The various in-in correlation functions are obtained by functionally differentiating $W[J_+,J_-]$
with respect to the sources and setting $J_+=J_-=0$ in the end.
Because there are now two types of fields and two types of sources,
there will be two kinds of vertices and four kinds of propagators involved
in Feynman diagrams, namely
\beq
G_{ab}(x,x')
=
\left.
\frac{\hbar\,\delta}{\text{sign}(a)\, i\,\delta J_a(x)}\,
\frac{\hbar\,\delta}{\text{sign}(b)\, i\,\delta J_b(x')}\,
e^{i\,W[J_+,J_-]}\right|_{J_+ = J_- = 0}
\ ,
\eeq
where
\beq
\text{sign}(a)
=
\begin{cases}
+1
\quad
{\rm for}\ 
a=+
\\
-1
\quad
{\rm for}\
a=-
\ .
\end{cases}
\eeq
The diagonal components of $G_{ab}$ correspond to the Feynman and anti-Feynman propagators,
\begin{align}
G_{++}(x,x')
& =
\bra{0_\text{in}} T\,\phi(x)\,\phi(x')\ket{0_\text{in}}
\\
G_{--}(x,x')
&
= 
\bra{0_\text{in}} \bar T\,\phi(x)\,\phi(x') \ket{0_\text{in}}
\ ,
\end{align}
where $T$ and $\bar T$ denote the time-ordered and anti time-ordered operators, respectively.
The off-diagonal components correspond to Wightman correlation functions,
\begin{align}
G_{+-}(x,x')
&
=
\bra{0_\text{in}} \phi(x')\,\phi(x) \ket{0_\text{in}}
\\
G_{-+}(x,x')
&
=
\bra{0_\text{in}} \phi(x)\,\phi(x') \ket{0_\text{in}}
\ .
\end{align}
Apart from the additional vertices and propagators, the in-in Feynman rules are identical
to the standard ones.
\par
For our purposes, the most important features of the Schwinger-Keldysh formalism are
the reality and causality of the in-in mean field $\bra{0_\text{in}}g_{\mu\nu}\ket{0_\text{in}}$,
and consequently of $\bra{0_\text{in}}\d s^2\ket{0_\text{in}}$.
These properties can be verified at every loop order by using the effective equations
derived from the in-in effective action $\Gamma[\phi_+,\phi_-]$, which is in turn given 
by the Legendre transform of the in-in generating functional $W[J_+,J_-]$ with respect
to the sources $J_\pm$.
The reality of the mean field is crucial for the interpretation of
$\bra{0_\text{in}}\d s^2 \ket{0_\text{in}}$ as a physical length,
whereas its causality uniquely determines the retarded Green's function
$G^\text{ret} = G_{++} - G_{+-}$ as the correct propagator to be used for the calculation
of the minimum length in the next section.
We refer the reader to Refs.~\cite{Jordan:1986ug,Barvinsky:1987uw} for the detailed
proof of the reality and causality of the mean field.
\section{Absence of a minimum length, presence of a minimum length scale}
\label{general}
In the present section, we use the results of Section~\ref{inin} to elaborate a model-independent
argument for the absence of a minimum geometrical distance to all orders of perturbation theory.
We only assume that the gravitational field is described by a metric tensor
$g_{\mu\nu}$ for which a background value $\bar g_{\mu\nu}$ exists in the
vacuum $\ket{0_{\rm in}}$, and on which its quantum fluctuations are free of interactions.
While the latter is obviously unrealistic, it should be enough for grasping the idea of a minimum length.
In fact, we would expect that a minimum length could exist as a consequence of quantum fluctuations,
which promote uncertainties in the proper length regardless of whether they are interacting or not.
\par
Instead of parameterizing the quantum field by the usual linear perturbation
$g_{\mu\nu} = \bar g_{\mu\nu} + h_{\mu\nu}$, we shall use the exponential parameterization
previously considered in Refs.~\cite{Nink:2014yya,Demmel:2015zfa,Nink:2015lmq},
that is~\footnote{With this parameterization, the quantum fluctuation $h_{\mu\nu}$ has
the dimensions of a canonical scalar field, that is $\sqrt{\rm mass/length}$.}
\begin{align}
g_{\mu\nu}
&
=
\bar g_{\mu\rho} \left(e^{\sqrt{\frac{32\,\pi\,\lp}{\mpl}}\,h}\right)^\rho_{\ \nu}
\nonumber
\\
&
=
\bar g_{\mu\nu}
+
\sqrt{\frac{32\,\pi\,\lp}{\mpl}}\,h_{\mu\nu}
+
\frac{16\,\pi\,\lp}{\mpl}\,h_{\mu\rho}\,h^{\rho}_{\ \nu}
+
O\left((\lp/\mpl)^{3/2}\right)
\ ,
\label{eq:param}
\end{align}
where $\lp = \sqrt{\gn\,\hbar}$ and $\mpl=\sqrt{\hbar/\gn}$ denote the Planck length and mass,
respectively.
The exponential parameterization has the advantage of transforming the problem of calculating
the expectation value of $\d s^2$ into the problem of computing correlation functions of the quantum
field $h_{\mu\nu}$.
Note that, classically, there is nothing that prevents the proper distance between two spacetime points
of coordinates $x^\mu$ and $y^\mu$ from going to zero in the limit in which $\d x^\mu=y^\mu-x^\mu$
vanish and the points coincide.
We thus expect 
\beq
\lim_{x\to y} \d s^2
=
\lim_{x\to y}
\left( \bar g_{\mu\nu}\, \d x^\mu\, \d x^\nu \right)
\equiv
\lim_{x\to y}
\left[
\ell^2(x,y)
\right]
=
0
\ ,
\label{class0}
\eeq
for any classical metric $\bar g_{\mu\nu}$.
Nonetheless, since the expectation value of quadratic and higher-order quantities evaluated at the
same spacetime event, such as $\bra{0_\text{in}}h_{\mu\rho}(x)\,h^{\rho}_{\ \nu}(x)\ket{0_\text{in}}$,
are divergent in quantum field theory, the coincidence limit of the quantum proper length must
be computed with care.
In fact, we must first regularize the divergences as there might be occasional cancelations leading
to a minimal length.
Because we are interested only in the coincidence limit, it is natural to isolate the divergences with
the covariant point-splitting, namely
\beq
\bra{0_\text{in}}h_{\mu\rho}(x)\,h^{\rho}_{\ \nu}(x)\ket{0_\text{in}}
=
\lim_{x\to y}\,
\bra{0_\text{in}}h_{\mu\rho}(x)\,h^{\rho}_{\ \nu}(y)\ket{0_\text{in}}
\ ,
\eeq
with similar expressions for higher-order correlators.
This allows us to write the quantum proper length in terms of correlation functions,
which at second order in $h_{\mu\nu}$ reads
\begin{align}
\lim_{x\to y}
\bra{0_\text{in}}\d s^2\ket{0_\text{in}}
&
=
\lim_{x\to y} 
\left(
\bra{0_\text{in}}g_{\mu\nu}\ket{0_\text{in}}\, \d x^\mu\, \d x^\nu
\right)
\nonumber
\\
&
=
\frac{16\,\pi\,\lp}{\mpl}\,
\lim_{x\to y}
\left[
\bra{0_\text{in}}h_{\mu\rho}(x)\,h^{\rho}_{\ \nu}(y)\ket{0_\text{in}}\,
\d x^\mu\, \d x^\nu
\right]
\nonumber
\\
&
\equiv
\frac{16\,\pi\,\lp}{\mpl}\,\lim_{x\to y}
\left[
G_{\mu\rho\ \ \mu}^{\ \ \ \rho}(x,y)\,\d x^\mu\, \d x^\nu
\right]
\ ,
\label{eq:ds}
\end{align}
where we used the expansion in Eq.~\eqref{eq:param} together with the fact that
the contribution at zero separation vanishes according to Eq.~\eqref{class0}, as well as does
the first order $\bra{0_\text{in}}h_{\mu\nu}\ket{0_\text{in}}=0$.
The question of a minimum length is thus translated into the calculation of the in-in
gravitational propagator $G_{\mu\rho\ \mu}^{\ \ \ \rho}$. 
But as we saw in Section~\ref{inin}, there are four different types of propagators
associated to in-in processes and, furthermore, they can be combined into other
propagators, such as the retarded and the advanced ones.
The immediate consequence is that $\bra{0_\text{in}}\d s^2\ket{0_\text{in}}$ 
appears ambiguous as there is a priori no reason to choose one propagator over the others.
In our case, the way we determine the relevant propagator should depend on how one measures
distances between two points at such (expectedly Planckian) small scales.
Such a measurement can take place via scattering processes (e.g.~to determine the mean free path),
which requires the Feynman propagator, or via the observation of a certain signal at different times
along its evolution, which would require the retarded Green's function.
Thus, the requirement of causality in the evolution of $\bra{0_\text{in}}\d s^2\ket{0_\text{in}}$
entails the use of the retarded Green's function~\cite{Jordan:1986ug,Barvinsky:1987uw}.
Note that mid-step calculations will involve all four types of Green's functions,
but the final result must necessarily depend solely on the retarded Green's function due to causality.
In fact, as shown in Refs.~\cite{Barvinsky:1987uw} (see also~\cite{Vilkovisky:2007ny}
for a detailed review), the in-in correlation functions (to any loop order) are obtained by
replacing form factors by retarded Green's functions.
In the asymptotically flat and empty spacetime, this agrees with the boundary conditions
in the remote past of the mean field.
\par
The calculation of propagators for an arbitrary curved background $\bar g_{\mu\nu}$
only add unnecessary complication, thus we shall take $\bar g_{\mu\nu} = \eta_{\mu\nu}$
as the Minkowski spacetime in the rest of this paper.
Our argument can then be generalised to curved spaces with the aid of the Schwinger
proper-time representation for propagators.
We shall also treat $h_{\mu\nu}$ as a free field and assume the gravitational propagator
in momentum space to take the simplest form of a sum over the number of simple poles
$m_i^2$ in the $q^2$-plane, that is
\beq
\Delta_{\mu\nu\rho\sigma}(q^2)
=
\sum_i \frac{\hbar\,P^i_{\mu\nu\rho\sigma}}{q^2-m^2_i}
\ ,
\label{eq:momprop}
\eeq
where
\beq
P^i_{\mu\nu\rho\sigma}
=
\alpha^i\, \eta_{\mu\rho}\,\eta_{\nu\sigma}
+ \beta^i \,\eta_{\mu\sigma}\,\eta_{\nu\rho}
+ \gamma^i\, \eta_{\mu\nu}\,\eta_{\rho\sigma}
\eeq
is the most general tensorial structure that can be combined into a tensor
of fourth rank and which is symmetric in $\{\mu\nu\}$ and $\{\rho\sigma\}$.
The coefficients $\alpha^i$, $\beta^i$ and $\gamma^i$ take different values
according to the theory at hand.
The propagator in position space is obtained from the $\epsilon$-prescription
or, equivalently, the integration contour corresponding to the retarded boundary
condition and reads
\beq
G^\text{ret}_{\mu\nu\rho\sigma}(x,y)
=
\sum_i \left[-\frac{\theta(x^0-y^0)}{2\,\pi}\,\delta(\ell^2)
+ \theta(x^0-y^0)\,\theta(\ell^2)\,
\frac{m_i\, J_1(m_i\,\ell)}{4\,\pi\,\ell}\right]
\hbar\,P^i_{\mu\nu\rho\sigma}
\ ,
\label{Gret}
\eeq
where $\ell^2\equiv \ell^2(x,y)=\eta_{\mu\nu}\,\d x^\mu\, \d x^\nu$ is the background proper distance
between $x$ and $y=x+\d x$.
The contraction $P_{\mu\rho\ \ \nu}^{i\ \ \rho}\,\d x^\mu\, \d x^\nu$ will always result
in a factor of $\ell^2$ in the numerator that can potentially be canceled by a divergence
$\ell^{-2}$ of the propagator, leaving a non-zero minimum length behind.
Note, however, that the first term above only contains a Dirac delta divergence
that cannot be canceled by $\ell^2$ and actually vanishes on integration,
whereas the second term diverges as $\ell^{-1}$ and cannot prevent $\ell^2$
from going to zero.
Putting this all together, gives
\beq
\lim_{x\to y}\,
\bra{0_\text{in}}\d s^2\ket{0_\text{in}}
=
\frac{16\,\pi\,\lp}{\mpl}\,
\lim_{x\to y} 
\left[
G^{\text{ret}\, \rho}_{\mu\rho\ \ \nu}(x,y)\,\d x^\mu\, \d x^\nu
\right]
= 0
\label{eq:zerolength}
\eeq
and we conclude that there is no minimum length to second order in $h_{\mu\nu}$.
\par
For an interacting theory, this does not imply the absence of a minimum length to all orders
in perturbation theory.
In the free theory, however, Wick's theorem can be used to reduce higher-order vacuum
correlation functions into a sum over products of the propagator, leading to
\beq
\expec{h^{n+2}}
\sim
\frac{1}{\ell^{n+2}}
\ ,
\qquad
n=1,2,\ldots
\ ,
\eeq
which suggests that there is no other relevant correlation function
(in addition to the one for $n=0$)
that could possibly cancel the vanishing length $\ell^2$ to produce a non-zero minimum length,
thus extending Eq.~\eqref{eq:zerolength} to all orders in $h_{\mu\nu}$.
This is in fact confirmed by the following non-perturbative calculation.
From Eq.~\eqref{eq:param} we have,
\begin{align}
\lim_{x\to y}\,
\bra{0_\text{in}}\d s^2\ket{0_\text{in}}
&=
\lim_{x\to y}\,
\left[\eta_{\mu\rho}\,
\bra{0_\text{in}}
\left(e^{\frac12\,\sqrt{\frac{32\pi\lp}{\mpl}}\,h(x)}\,
e^{\frac12\,\sqrt{\frac{32\pi\lp}{\mpl}}\,h(y)} \right)^\rho_{\ \nu}
\ket{0_\text{in}}\,\d x^\mu\, \d x^\nu\right]
\nonumber
\\
&=
\lim_{x\to y}\,
\left[\eta_{\mu\rho}\left(e^{\frac{8\pi\lp}{\mpl}\,\bra{0_\text{in}}\,h(x)\,h(y)\,\ket{0_\text{in}}}\right)^\rho_{\ \nu}\,
\d x^\mu\, \d x^\nu\right]
\nonumber
\\
&=
\lim_{x\to y}\,
\left[\ell^2 \,e^{-{4\,\lp^2}\,\theta(x^0-y^0)\,\delta(\ell^2)\, \sum_i(\alpha^i+4\beta^i+\gamma^i)}\right]
\nonumber
\\
&= 0,
\end{align}
where we used point-splitting in the first line, applying normal ordering in both exponential operators separately,
and the Baker-Campbell-Hausdorff formula together with Wick's theorem in the second equality.
The third equality is obtained by manipulating the exponential as an infinite series and resumming
back to the exponential form~\footnote{We defined the product of Dirac deltas as a convolution
$\delta^2\to\delta * \delta = \delta$.}.
Free gravitational fluctuations are thus not prone to minimum length.
Even when interactions are switched on, loop corrections to the free propagator
cannot change this picture at second order.
In fact, the dressed propagator can be written in the K\"all\'en-Lehmann spectral
representation in terms of the free propagator itself as
\beq
G^\text{dressed}_{\mu\nu\rho\sigma}(x,y)
=
\int_0^\infty \d\mu^2 \,\rho(\mu^2)\,
G^\text{ret}_{\mu\nu\rho\sigma}(x-y;\mu^2)
\ ,
\eeq
where $\rho(\mu^2)$ is the spectral density.
Therefore, replacing $G^\text{ret}$ with $G^\text{dressed}$ in
Eq.~\eqref{eq:zerolength} would still give zero.
However, in the interacting theory one can no longer rely on Wick's theorem
to express higher-order correlation functions as products of the two-point function.
The vanishing of $\bra{0_\text{in}}\d s^2\ket{0_\text{in}}$ at second order
does therefore not allow us to come to any definite conclusion about the
existence of a minimum length in an interacting theory.
\par
Before continuing, let us comment on the in-out proper ``length''
$\bra{0_\text{out}}\d s^2\ket{0_\text{in}}$.
Although we have emphasized that it cannot be interpreted as a physical length
or a statistical quantity, it might suggest the existence of a minimum length
\textit{scale}.
If we repeat the above argument for the in-out amplitude,
we find
\begin{align}
\lim_{x\to y}\,
\bra{0_\text{out}}\d s^2\ket{0_\text{in}}
&
=
\mathcal{N}\lim_{x\to y} 
\left[
\ell^2\, e^{\frac{2\lp^2}{\pi\ell^2}\,\sum_i(\alpha^i+4\beta^i+\gamma^i)}
\right]
\nonumber
\\
&=
\begin{cases}
 \frac{2\,\lp^2}{\pi}\,
\sum_i 
\left(\alpha^i + 4\,\beta^i + \gamma^i
\right)
\sim
\lp^2
&
{\rm for}
\
\sum_i(\alpha^i+4\beta^i+\gamma^i) > 0
\\
0
&
{\rm for}
\
\sum_i(\alpha^i+4\beta^i+\gamma^i) \leq 0
\ ,
\end{cases}
\label{eq:scale}
\end{align}
where $\mathcal{N} = \langle 0_\text{out} | 0_\text{in}\rangle$ is a normalization factor chosen
to cancel divergences at $\ell = 0$. We used the Feynman propagator for small distances,
\begin{equation}
G^\text{F}_{\mu\nu\rho\sigma}(x,y)
=
\sum_i \frac{\hbar\,P^i_{\mu\nu\rho\sigma}}{4\,\pi^2\left(x-y\right)^2}
+ \mathcal O(|x-y|)
\ ,
\label{eq:propGR}
\end{equation}
which is obtained by Fourier transforming Eq.~\eqref{eq:momprop} with Feynman
boundary conditions.
Note that the exponential is divergent, but one can isolate the divergences
from the finite part by expanding the exponential function as a Taylor series.
The zeroth order term is simply $\ell^2$ which goes smoothly to zero.
The first order term in the expansion is also finite (but non-zero) because
of the cancelation of $\ell^2$ in the numerator and in the denominator.
Higher-order terms contain the divergences.
Nonetheless, one can use the arbitrariness of the normalization factor $\mathcal N$
to cancel the divergent part so that the limit $\ell\to 0$ is finite.
Eq.~\eqref{eq:scale} points at the
Planck scale as a potential limiting factor that screens everything that goes beyond it.
This is not to say that physical distances cannot vanish, but it suggests that scattering
experiments cannot tell apart trans-Planckian effects.
In the foreseeable future, astrophysics and cosmology seem to be the only hope
to probe quantum gravity experimentally.
\par
We kept the argument completely general, without the need of specifying the
gravitational theory, thus the conclusions above are quite general with the only
restriction that the gravitational field be described solely in terms of the metric.
Different theories will differ by their propagators with different values for the
coefficients $\alpha^i$, $\beta^i$ and $\gamma^i$, but they will all produce
vanishing minimum lengths and non-zero minimum length scales of Planckian
order unless
\beq
\sum_i(\alpha^i+4\beta^i+\gamma^i) \leq 0
\ .
\eeq
In general relativity, for example, the massless spin-2 field (graviton)
is the only degree of freedom,
\beq
\hbar^{-1}
\Delta_{\mu\nu\rho\sigma}
=
\frac{\eta_{\rho\mu}\,\eta_{\sigma\nu}+\eta_{\sigma\mu}\,\eta_{\rho\nu}-\eta_{\mu\nu}\,\eta_{\rho\sigma}}
{q^2}
\ .
\eeq
The above considerations imply that no minimum length exists for general relativity,
but a minimum length scale is again inferred from
\beq
\lim_{x\to y}\,
\bra{0_\text{out}}\d s^2\ket{0_\text{in}}
=
\frac{8\,\lp^2}{\pi}
\ .
\label{Lscale}
\eeq
More general theories of gravity are expected to contain other degrees of freedom
in addition to the graviton.
This is evident in higher-derivative theories where new degrees of freedom are
essential for the renormalizability of the theory.
For example, the propagator of Stelle's theory reads~\cite{Stelle:1977ry,Stelle:1976gc}
\beq
\hbar^{-1}
\Delta_{\mu\nu\rho\sigma}
=
\frac{2\,P^{(2)}_{\mu\nu\rho\sigma}- P^{(0)}_{\mu\nu\rho\sigma}}{q^2}
-\frac{2\, P^{(2)}_{\mu\nu\rho\sigma}}{q^2-m_2^2}
+ \frac{ P^{(0)}_{\mu\nu\rho\sigma}}{q^2-m_0^2}
\ ,
\label{eq:hdprop}
\eeq
where $P^{(i)}_{\mu\nu\rho\sigma}$ are spin-projection operators,
and one can see the additional massive degrees of freedom,
namely a scalar excitation of mass $\hbar\,m_0$ and a spin-2 particle
of mass $\hbar\,m_2$, that turn out to make the theory renormalizable.
The minimum length scale in this case vanishes
\beq
\lim_{x\to y}\,
\bra{0_\text{out}}\d s^2\ket{0_\text{in}}
=
0
\ .
\label{eq:cancelations}
\eeq
due to accidental cancelations of the coefficients in the numerator $\sum_i \left(\alpha^i + 4\,\beta^i + \gamma^i \right) = 0$.
When self-interactions are considered for $h_{\mu\nu}$, all the three degrees
of freedom will couple to each other, making the whole analysis much more difficult.
In this scenario, Wick's theorem is of no help to us and nothing can be said
about the contributions from higher-order correlation functions,
thus a non-perturbative treatment is certainly desirable.
This is the subject of the following Section.
\section{A non-perturbative example: higher-derivative gravity}
\label{hdgravity}
In this section, we compute the quantum proper length
$\bra{0_\text{in}}\d s^2\ket{0_\text{in}}$ non-perturbatively
for higher-derivative gravity without resorting on the exponential
parameterization used in the last Section.
The idea is to perform field redefinitions in the action in order to make
the additional degrees of freedom explicit from the outset. 
\par
The action of higher-derivative gravity reads
\begin{align}
\label{eq:localaction}
S
=
\frac{\mpl}{16\,\pi\,\lp}
\int \d^4x\,
\sqrt{- g}
\left(R + c_1\, R^2 + c_2\, R_{\mu\nu} \,R^{\mu\nu}
+ c_3\, R_{\mu\nu\rho\sigma}\, R^{\mu\nu\rho\sigma}
\right)
\ ,
\end{align}
where $R$, $R_{\mu\nu}$ and $R_{\mu\nu\rho\sigma}$ are the Ricci scalar,
Ricci tensor and Riemann tensor of the metric $g_{\mu\nu}$,
respectively,~\footnote{Note that the square of the Riemann
tensor is usually eliminated in favour of the other two curvature
invariants by invoking Gauss-Bonnet theorem.
Here we choose to leave it explicit in the action just to follow the same
notations commonly used in the literature.}
and $c_i$ are dimensionful coupling constants.
The above action contains massive particles of spin-0 and spin-2
in addition to the usual graviton which corresponds to the massless
spin-2 excitation.
All these degrees of freedom can be made explicit in the action
via a Legendre transform~\cite{Magnano:1990qu} followed
by a field redefinition of the form~\cite{Hindawi:1995an}
\begin{equation}
g_{\mu\nu}
=
e^{-\sqrt{\frac{16\pi\,\lp}{3\,\mpl}}\,\chi}
\,\bar g_{\mu\nu}
\ ,
\label{eq:map}
\end{equation}
resulting in the action~\cite{Hindawi:1995an}
\begin{align}
S
=
\int \d^4 x\,\sqrt{-\bar g}
&
\left[
\frac{\mpl}{16\,\pi\,\lp}\,\bar R
-\frac{1}{2}\,\bar \nabla^\mu \chi\, \bar \nabla_\mu \chi
-\frac{3\,\mpl}{32\,\pi\,\lp}\,m_0^2
\left(1-e^{-\sqrt{\frac{16\pi\,\lp}{3\,\mpl}}\,\chi}\right)^2
\right.
\nonumber
\\
&
\left.
\quad
-\frac{\mpl}{16\,\pi\,\lp}\bar G_{\mu\nu}\,\pi^{\mu\nu}
+ \frac{\mpl}{64\,\pi\,\lp}\,m_2^2 \left(\pi_{\mu\nu}\, \pi^{\mu\nu}-\pi^2\right)\right]
\ ,
\label{eq:newframeaction}
\end{align}
where $\pi\equiv \bar g^{\mu\nu}\,\pi_{\mu\nu}$, 
$m_0 = (6\,c_1 + 2\,c_2 + 2\,c_3)^{-1/2}$ is the inverse Compton length
of the scalar field $\chi$ and $m_2=(-c_2-4\,c_3)^{-1/2}$ that of the
massive spin-2 particle $\pi_{\mu\nu}$.
Note that the action for $\pi_{\mu\nu}$ is not in canonical form
(it does not even contain a kinetic term).
Canonicalizing $\pi_{\mu\nu}$ requires an additional field redefinition
(see Ref.~\cite{Hindawi:1995an}).
This additional field redefinition gives rise to the kinetic term of $\pi_{\mu\nu}$
as well as it makes explicit the coupling between $\pi_{\mu\nu}$ and $\chi$.
Nonetheless, the frame with a canonical $\pi_{\mu\nu}$ is no better than any other frame.
We chose to work in the frame~\eqref{eq:newframeaction} because it simplifies
the calculation of the minimum length.
\par
We interpret $\bar g_{\mu\nu}$ as a classical background~\footnote{Fluctuations
$h_{\mu\nu}$ of $\bar g_{\mu\nu}$ would only contribute to higher-order terms
$h\,\chi\,\Box\chi \sim \mathcal O(E^5/\mpl^5)$, which are negligible to leading order.}
where the quantum fields $\chi$ and $\pi_{\mu\nu}$ live on and, as before,
we consider the Minkowski background $\bar g_{\mu\nu} = \eta_{\mu\nu}$.
Since there is no explicit interaction of $\chi$ with $\pi_{\mu\nu}$ in the
action~\eqref{eq:newframeaction}, we can focus solely on the spin-0 sector.
From the translational symmetry of the path integral measure,
we can shift $\chi\to\chi+\chi_0$ and take $\chi_0\to\infty$,
which simplifies the spin-0 action to \cite{Cunliff:2012zb}
\beq
S_\chi
=
\frac{1}{2}\int \d^4 x\,  \chi\,\Box \chi
\ ,
\label{eq:scalar}
\eeq
where we discarded a constant term as it does not contribute to the
equations of motion.
The retarded propagator for $\chi$ is thus simply given by the propagator of a massless scalar field \cite{Cunliff:2012zb}
\beq
\bra{0_\text{in}}\chi(x)\,\chi(y)\ket{0_\text{in}}
=
-\hbar\,
\frac{\theta(x^0-y^0)}{2\,\pi}\,\delta(\ell^2)
\ .
\label{eq:scalarprop}
\eeq
From Eqs.~\eqref{eq:map} and~\eqref{eq:scalarprop},
the quantum proper length in the in-vacuum state vanishes in the
coincidence limit as
\begin{align}
\lim_{x\to y}\,
\bra{0_\text{in}}\d s^2\ket{0_\text{in}}
&
=
\lim_{x\to y}
\left[
\bra{0_\text{in}}
e^{-\frac12\sqrt{\frac{16\pi\,\lp}{3\,\mpl}}\,\chi(x)}
\,e^{-\frac12\sqrt{\frac{16\pi\,\lp}{3\,\mpl}}\,\chi(y)}
\ket{0_\text{in}}\,
\eta_{\mu\nu}\,\d x^\mu\, \d x^\nu
\right]
\nonumber
\\
&
=
\lim_{x\to y}
\left[
\ell^2 \, e^{\frac{4\pi\lp}{3\,\mpl}\,\bra{0_\text{in}}\chi(x)\,\chi(y)\ket{0_\text{in}}}
\right]
\nonumber
\\
&
=
0
.
\label{eq:nonpert}
\end{align}
As before, we performed a point-splitting in the first line, imposing normal ordering
in each of the exponential operators separately.
The second equality follows from the Baker-Campbell-Hausdorff formula in combination
with Wick's theorem~\footnote{Notice that we started with the full interacting theory
Eq.~\eqref{eq:newframeaction}, but we managed to reduce the scalar sector to that of
a free scalar field~\eqref{eq:scalar}, which permits the application of the Wick's theorem.
That is not to say that $\chi$ is physically free of interactions as the action
we started with does contain interaction terms among all degrees of freedom.
In fact, choosing a background other than Minkowski would invalidate our argument,
since the path integral measure would no longer have translational symmetry.
Non-trivial path integral contributions would then come into play, making the calculation
very difficult at the non-perturbative level.
As long as we make the simplifying assumption that the background is Minkowski,
non-perturbative calculations are possible.
The same applies to the example of Sec.~\ref{conf}}..
Therefore, the finding~\eqref{eq:nonpert} confirms that the vanishing
of the quantum proper length observed in Eq.~\eqref{eq:zerolength}
for non-interacting fluctuations $h_{\mu\nu}$ actually extends to the
interacting case as well.
Similarly, the in-out proper ``length'' reads
\begin{align}
\lim_{x\to y}\,
\bra{0_\text{out}}\d s^2\ket{0_\text{in}}
&
=
\mathcal{N}\,\ell^2\,
\lim_{x\to y}
\left[
e^{\frac{4\pi\lp}{3\,\mpl}\,\bra{0_\text{out}}\chi(x)\,\chi(y)\ket{0_\text{in}}}
\right]
\nonumber
\\
&
=
\frac{\lp^2}{3\,\pi}
\ ,
\label{eq:nonpert-inout}
\end{align}
where we again chose the normalization factor $\mathcal{N}$ to absorb
the divergence and we used
\beq
\bra{0_\text{out}}\chi(x)\,\chi(y)\ket{0_\text{in}}
=
\frac{\hbar}{4\,\pi^2\, (x-y)^2}
\ .
\eeq
This shows that the finite part of $\bra{0_\text{out}}\d s^2\ket{0_\text{in}}$
is not zero and indicates the existence of a minimum length scale.
It is important to stress that Eq.~\eqref{eq:nonpert-inout} is a non-perturbative
result which takes into account all interactions between the degrees of freedom
present in the theory.
This explains the difference with respect to the non-interacting case in
Eq.~\eqref{eq:cancelations}.
\section{Revisiting the conformal degree of freedom}
\label{conf}
In Ref.~\cite{Padmanabhan:1985jq}, it was argued that a Planckian minimum
length exists when one quantizes the conformal degree of freedom of
general relativity on a classical background.
This was performed by first parameterizing the metric as~\footnote{We keep
the field $\phi$ dimensionless here, instead of choosing the canonical normalization
of previous Sections, in order to ease the comparison with the
original work~\cite{Padmanabhan:1985jq}.}
\beq
g_{\mu\nu}
=
\left(1+\phi\right)^2
\bar g_{\mu\nu}
\ ,
\label{eq:conf}
\eeq
which separates the conformal degree of freedom $\phi$ from the other
degrees of freedom present in the classical background $\bar g_{\mu\nu}$.
In the parameterization~\eqref{eq:conf}, the Einstein-Hilbert action becomes
\beq
S
=
\frac{\mpl}{16\,\pi\,\lp}\int\d^4x\,\sqrt{-\bar g}
\left[\bar R
\left(1+\phi\right)^2
-2\,\Lambda\left(1+\phi\right)^4
- 6\,\partial_\mu\phi\,\partial^\mu\phi\right]
\ .
\eeq
In a Minkowski background, namely $\bar R = \Lambda = 0$,
the action effectively becomes that of a free and massless scalar field.
Because of the simplicity of the action when $\bar g_{\mu\nu} = \eta_{\mu\nu}$,
one is able to perform non-perturbative calculations.
Upon quantizing the conformal degree of freedom $\phi$, its Feynman propagator
can be easily obtained as~\footnote{The non-standard numerical factor appears
because of the non-canonical normalization of $\phi$.}
\beq
\bra{0_\text{out}}\phi(x)\,\phi(y)\ket{0_\text{in}}
=
\frac{\hbar\,\lp^2}{3\,\pi \, \mpl \,(x-y)^2}
\ .
\label{eq:feynman}
\eeq
The quantum proper distance $\bra{0_\text{out}}\d s^2\ket{0_\text{in}}$
in the in-out formalism was then calculated with the aid of the point-splitting
regularization as in Section~\ref{general}.
One therefore obtains 
\begin{align}
\lim_{x\to y}\,
\bra{0_\text{out}}\d s^2\ket{0_\text{in}}
&
=
\lim_{x\to y}
\left[
\bra{0_\text{out}}\phi(x)\,\phi(y)\ket{0_\text{in}}\,
\eta_{\mu\nu}\,\d x^\mu\, \d x^\nu
\right]
\ ,
\nonumber
\\
&
=
\frac{\lp^2}{3\,\pi}
\ ,
\label{eq:inoutconf}
\end{align}
which precisely equals the result~\eqref{eq:nonpert-inout}.
\par
However, as we stressed previously, $\bra{0_\text{out}}\d s^2\ket{0_\text{in}}$
should not be interpreted as a physical distance because it is a complex number
in general.
Eq.~\eqref{eq:inoutconf} only gives a real result because it was computed at
the tree level, but when loop corrections are taken into account, an imaginary
part shows up in Eq.~\eqref{eq:inoutconf}.
The correct way of computing geometrical distances at the quantum level is
via in-in amplitudes, in which case we must replace the Feynman
propagator~\eqref{eq:feynman} with the retarded propagator~\eqref{eq:scalarprop} 
(with $\phi$ in place of $\chi$ and taking into account the field normalizations),
which yields
\begin{align}
\lim_{x\to y}\,
\bra{0_\text{in}}\d s^2\ket{0_\text{in}}
&
=
\lim_{x\to y}\,
\left[\left(1 
+\bra{0_\text{in}}\phi(x)\,\phi(y)\ket{0_\text{in}}
\right)
\ell^2
\right]
\nonumber
\\
&= 0,
\end{align}
showing, once again, the absence of a minimum length.
\section{Conclusions}
\label{conc}
In this paper, we have reconsidered the idea of a minimum
geometrical length in quantum gravity through the lens
of the Schwinger-Keldysh formalism, from which in-in amplitudes
can be derived.
Because the in-in quantum proper distance is calculated from a single state,
one is able to interpret it as a truly geometrical length that
happens to be real at all loop orders and satisfies a causal equation
of motion, which is manifested via retarded Green's functions.
When the in-in proper length is evaluated at coinciding points,
we used pertubative arguments to show it vanishes at second order
for any metric theory of gravity.
In the absence of interactions, this result can be extended to all
orders of perturbation theory.
Under suitable reparametrizations of the metric, we also showed
non-perturbatively that a minimum length cannot exist in higher-derivative
gravity or in the conformal sector of general relativity.
Whereas the requirement of reality should be obvious for the notion
of a geometrical distance, one might argue why causality is also a 
welcome property.
The use of the retarded propagator demanded by the in-in formalism
implies that quantum corrections to the distance between two spacetime
points will always vanish when the points lie outside the respective
light cones in the background metric.
This result therefore appears as a consistency condition for the very
existence of a background metric and the geometrical description of
gravity.~\footnote{Note that the background metric
could still be determined self-consistently by solving effective field equations
which include loop corrections without affecting our argument.}
Moreover, and indeed equivalently, this result implies that the free
propagation of physical signals of any frequency will not be affected by a fundamental
length scale.
Their dispersion relation will be simply determined by the background metric
and quantum gravity effects cannot be probed by detecting the way signals
travel through spacetime.
\par
While a geometrical minimum length seems to be unlikely,
we made the case for a minimum length scale, namely the scale extracted
from the in-out amplitude $\bra{0_\text{out}}\d s^2\ket{0_\text{in}}$
at the coincidence limit.
By following the same reasoning as for the in-in length, we found theoretical
evidence that points at the Planck length as a universal scale beyond
which scattering experiments become useless as, even in principle,
they cannot distinguish between physical effects taking place at
energies $E\gtrsim\mpl$.
This only reinforces the need for a change of paradigm in quantum field theory
from scattering experiments to time-dependent evolutions, which signifies
the importance of in-in amplitudes in physics.
Of course, one could further argue that most physical processes involve
scatterings at some level.
For instance, the physical signals we can detect will have been produced
by interactions, whose field theoretic description is given in terms of an $S$-matrix
involving Feynman propagators.
Here is where the minimum length \textit{scale} seems to enter the picture
again, opening up the possibility of probing quantum gravity indirectly
from the imprints left in the signals at lower energies.~\footnote{We
also mention that scatterings at trans-Planckian
energies are expected to produce (quantum or semiclassical) black holes and
the issue of a minimum length becomes intertwined with the features of such
non-perturbative configurations, which generalised uncertainty principles were
envisioned to account for~\cite{Scardigli:1999jh,Hossenfelder:2012jw}.}
\par
We would like to conclude by remarking once more that the basic assumption
in our analysis is the existence of a background metric (irrespectively of what
that metric actually is).
Approaches which lead to the appearance of a minimum geometric length 
must somehow violate this requirement.
For instance, the resemblance of general relativity to thermodynamics~\cite{ted}
suggests that the classical geometry of spacetime is an emergent phenomenon,
very much like the notion of thermodynamics for a classical fluid emerges from
the statistical mechanics of a more fundamental microscopic theory.
Waves in such a fluid can be produced and freely propagate only if their
wavelength is significantly larger than the scale of the underlying microscopic
structure.
This brings forth the questions of what is the fundamental dynamics of
gravity at the Planck scale and, not less important, what is the quantum state
$\ket{0_{\rm in}}$, which describe the Universe as we see it.
Results from effective field theoretic descriptions at experimentally accessible
scales can hopefully serve as a guideline in this quest.
\section*{Acknowledgments}
I.K.~and R.C.~are partially supported by the INFN grant FLAG.
The work of R.C.~has also been carried out in the framework of activities
of the National Group of Mathematical Physics
(GNFM, INdAM) and COST action Cantata.
%
%
%
%

%
\end{document}